\documentclass[conference]{IEEEtran} 
\IEEEoverridecommandlockouts

\usepackage{cite}
\usepackage{amsmath,amssymb}
\usepackage{graphicx}
\usepackage{float} 
\usepackage{tabularx}
\usepackage{booktabs}
\usepackage{multirow}
\usepackage{url}
\usepackage[hidelinks]{hyperref}
\usepackage{iftex}
\ifXeTeX 
  \usepackage{fontspec}
\else
  \usepackage[T1]{fontenc} 
\fi
\newcommand{\nBudgetMaxMeanAUROC}{0.962}

\newcommand{\nBudgetModelsWithHiddenFamilyNoSAA}{9}

\newcommand{\nBudgetNoSAAHiddenZulu}{8}

\newcommand{\nBudgetTopMeanTPRone}{0.416}

\newcommand{\nCalibLtwoNanoDolly}{0.929}

\newcommand{\nCalibMTKFPRratioMax}{5.4}
\newcommand{\nCalibMTKFPRratioMin}{2.3}
\newcommand{\nCalibMTKFPRrealMax}{0.272}
\newcommand{\nCalibMTKFPRrealMin}{0.116}

\newcommand{\nIdentityMaxDiff}{6.1\times10^{-5}}

\newcommand{\nLthreeSAAAUROC}{0.918}

\newcommand{\nLthreeSAATPR}{0.002}

\newcommand{\nLthreeSAAeightTPR}{0.007}

\newcommand{\nLtwoFPR}{0.054}
\newcommand{\nLtwoFPRsevenfive}{0.076}

\newcommand{\nLtwoMarkupSpearman}{0.83}

\newcommand{\nLtwoSAAAUROC}{0.942}

\newcommand{\nLtwoSAATPR}{0.114}

\newcommand{\nLtwoSAATPRsevenfive}{1.000}

\newcommand{\nLtwoSAAeightTPR}{0.112}

\newcommand{\nMisBankOnlySAATPRmax}{1.000}
\newcommand{\nMisBankOnlySAATPRmin}{0.000}

\newcommand{\nMisLotterySAATPRmax}{1.000}
\newcommand{\nMisLotterySAATPRmed}{0.303}
\newcommand{\nMisLotterySAATPRmin}{0.000}

\newcommand{\nVicSAAeightTPR}{0.018}

\newcommand{\nWCDollyAbove}{19}

\newcommand{\nWCPairs}{20}

\usepackage{soul}
\usepackage{xcolor}

\begin{document}

\title{Beyond AUROC: Deployment-Aware Evaluation of Model-Internal Jailbreak Detectors}

\author{\IEEEauthorblockN{Mohammad Tehabsim\IEEEauthorrefmark{1}, Basil Al-Housani\IEEEauthorrefmark{1}, Nour Alhussien, and Phung Lai}
\IEEEauthorblockA{Department of Cybersecurity\\ University at Albany, SUNY\\ Albany, NY, USA\\{mtehabsim, balhousani, nalhussien, lai}@albany.edu\\
\IEEEauthorrefmark{1}Equal contribution}}

\maketitle

\begin{abstract}
Large language models (LLMs) are increasingly deployed in user-facing systems, where jailbreak attacks can bypass safety controls and induce harmful and policy-violating outputs. Recent model-internal detectors use hidden representations, gradients, or internal signals to identify such prompts, but their practical security value depends critically on behavior at low false-positive operating points rather than on area under the receiver operating characteristic curve (AUROC) alone.

We study two data dependencies that shape this behavior: \emph{calibration data}, which determine the detection threshold, and \emph{reference data}, which can influence detector scores. Across four model-internal detectors, four LLMs, and twelve jailbreak families, thresholds calibrated on curated benign prompts frequently fail to preserve their false-positive rate (FPR) budgets on real-user traffic. Recalibration improves false-positive control but reveals substantial family-level coverage gaps, including cases with high AUROC but low true-positive rate (TPR). Causal reference-set interventions show that changing reference composition can substantially alter family-level detection and that improving coverage of one attack family can reduce coverage of others.

These findings show that calibration and reference data are part of a detector's security behavior. We therefore recommend evaluating model-internal jailbreak defenses using realized FPR on deployment-like traffic and per-family TPR at explicit low-FPR operating points, with AUROC used as a complementary ranking metric rather than the primary security measure.
\end{abstract}

\begin{IEEEkeywords}
jailbreak detection, large language models, Model-Internal Representations, calibration, Security Evaluation
\end{IEEEkeywords}

\section{Introduction}
Large language models (LLMs) increasingly rely on alignment mechanisms to refuse harmful or policy-violating requests \cite{xu2024dpo,zhou2025don}. Jailbreak attacks challenge these safeguards by inducing aligned models to produce content they would otherwise refuse \cite{qi2024visual}, potentially exposing users to harmful outputs and weakening deployed safety controls. As jailbreak strategies have become more diverse, recent defenses increasingly inspect internal model signals, including hidden representations, gradients, and refusal-related directions \cite{mtk,gradsafe,hiddendetect}. We refer to these approaches as \emph{model-internal jailbreak detectors}. Rather than relying solely on surface-level prompt features, these detectors seek evidence of malicious intent within the model's internal computation.

Despite this progress, three deployment-relevant aspects are not jointly characterized in current evaluations. First, thresholds calibrated on curated benign datasets may not preserve the same false-positive rate (FPR) on deployment-like traffic, creating a \emph{calibration-transfer gap}. Second, detectors such as MTK \cite{mtk}, GradSafe \cite{gradsafe}, and supervised activation probes \cite{goldowsky} depend on \emph{reference data} that can influence detector scores and thresholded decisions. Evaluations based on a single fixed reference configuration may therefore obscure the sensitivity to reference composition, creating a \emph{reference-sensitivity gap}. Third, model-internal detectors are commonly summarized using threshold-independent metrics such as AUROC or AUPRC \cite{mtk,hiddendetect,gradsafe}. These metrics characterize global ranking quality but do not directly measure attack coverage at the low FPRs required in deployment. A detector may therefore achieve high AUROC while detecting only a small fraction of a particular attack family at the deployed threshold. Although this distinction is well established in statistical detection and security evaluation \cite{bradley,mcclish,carlinimia}, it creates an \emph{operating-point gap} between aggregate metrics and practical family-level coverage.

These gaps are coupled through the detector's decision process. Reference data can alter the score assigned to a prompt, while benign calibration data determine the threshold used to convert that score into a security decision. As a result, the same detector architecture can exhibit different family-level behavior under different reference or calibration data. Treating these data choices as incidental experimental details can therefore hide security-relevant variability at the deployed operating point.

This paper studies model-internal jailbreak detection as a data-dependent decision problem. Our evaluations cover detectors based on manifold trajectories, gradient similarity, refusal-related representations, and supervised activation probes, together with a surface-level perplexity baseline. We evaluate them under a common framework that separates calibration effects from reference-data effects, measure detector behavior across multiple jailbreak families and LLMs, and use controlled causal reference interventions to determine whether changes in reference composition directly alter family-level coverage.

Our contributions are as follows:
\begin{itemize}
   \item We formulate model-internal jailbreak detection as a data-dependent decision process in which \emph{reference data} influence detector scores and \emph{calibration data} determine the operating threshold. This view makes two often implicit sources of deployment variability explicit.

    \item We characterize \emph{calibration transfer} by measuring realized FPR and attack-family detection when thresholds calibrated on curated benign prompts are applied to deployment-like user distributions.

    \item We evaluate \emph{reference sensitivity} through controlled reference-set interventions that isolate how changes in reference composition affect score construction, ranking behavior, and family-level coverage.

    \item We evaluate model-internal detectors at explicit low false-positive operating points and report per-family detection alongside aggregate ranking metrics, exposing failure modes that cannot be characterized by AUROC or AUPRC alone.

    \item We derive practical evaluation recommendations that emphasize deployment-like calibration, per-family operating points, sensitivity across reference draws, and explicit auditing of reference data.
\end{itemize}

The remainder of this paper is organized as follows. Section~\ref{sec:bg} reviews related work, Section~\ref{sec:method} introduces our data-dependent evaluation framework and formalizes the roles of reference data, calibration data, and operating thresholds in detector behavior. Section~\ref{sec:setup} describes the experimental setup, and Section~\ref{sec:res} presents the experimental results. Section~\ref{sec:disc} discusses the implications and limitations, and section~\ref{sec:conc} concludes the paper.

\section{Background and Related Work}
\label{sec:bg}

\subsection{Jailbreak Attacks and Model-Internal Defenses}
Jailbreak attacks aim to bypass the safety alignment of LLMs and induce harmful or unsafe responses that the model would otherwise refuse. Existing attacks use diverse strategies, including adversarial suffix optimization \cite{advbench,saa}, rewriting and iterative refinement \cite{autodan,pair}, persuasive prompting \cite{pap}, and multilingual transformations \cite{zulu}. This diversity complicates detection because similar harmful intent can appear through substantially different prompt surface forms.

Surface-level defenses such as perplexity filtering \cite{alon,jain} inspect the input text and can detect attacks that introduce unusual token patterns. However, many jailbreaks preserve harmful intent while changing the prompt's wording and structure, limiting the reliability of purely surface-level cues. Model-internal jailbreak detection addresses this limitation by examining signals produced during an LLM's internal computation, such as hidden states, gradients, or refusal-related representations \cite{mtk,gradsafe,hiddendetect}. These signals may retain information about malicious intent even when the prompt is reformulated at the surface level. Representative methods include MTK \cite{mtk}, which models layer-wise relationships to benign and malicious anchors; GradSafe \cite{gradsafe}, which compares safety-related gradients; HiddenDetect \cite{hiddendetect}, which analyzes hidden states; and supervised activation probes \cite{goldowsky,ccpp}. This motivates our focus on model-internal detectors, while also raising an important evaluation question: their behavior may depend not only on the query itself, but also on the reference and calibration data used to construct scores and decision thresholds.
 
\subsection{Reference Dependence in Model-Internal Detection}

Several model-internal detectors rely on reference data as part of their scoring mechanism. MTK uses benign and malicious anchors to construct layer-wise rank trajectories, GradSafe compares queries against safe and unsafe gradient references, and supervised probes learn decision boundaries from labeled internal representations \cite{mtk,gradsafe,goldowsky,ccpp}. Consequently, the detector score is determined not only by the query, but also by the reference data used to instantiate the detector. Changes in reference composition can therefore alter score rankings and, after thresholding, the coverage of individual attack families. Yet evaluations commonly report results for a fixed reference configuration, leaving unclear how sensitive the reported performance is to this choice. This motivates our analysis of reference sensitivity across reference draws and causal interventions in reference composition.

\subsection{Security Evaluation and Operating Points}
Most detectors output a continuous score that must be converted into a binary decision using a threshold. At a given threshold, the FPR measures benign prompts incorrectly flagged, while the true-positive rate (TPR) measures attacks detected. In security settings, this trade-off is operationally critical because even modest FPRs can be costly when benign inputs dominate \cite{axelsson,sommer,arp,tesseract,transcend}.

Model-internal jailbreak detectors are often summarized using threshold-independent metrics such as AUROC and AUPRC \cite{mtk,gradsafe,hiddendetect}. These metrics characterize overall ranking quality but do not determine how much of a specific attack family is detected at a deployment threshold. A detector can therefore achieve high AUROC while missing most prompts from an attack family at a strict false-positive budget. Carlini et al.~\cite{carlinimia} similarly argue in the security setting that AUROC can obscure practically relevant behavior because it averages performance over the full false-positive range, including high-FPR regions that may be operationally irrelevant. They advocate reporting TPR at low FPRs instead. Prior work has also emphasized fixed-FPR measures and partial AUROC for evaluating security-relevant operating regions \cite{bradley,mcclish}.

 \subsection{Calibration and Distribution Shift}
Operating thresholds are typically selected using benign calibration data to satisfy a target false-positive budget. However, that operating point is tied to the calibration distribution: when benign traffic changes, the same threshold may no longer preserve the intended FPR. This problem is well established in work on dataset shift, out-of-distribution detection, and conformal inference \cite{bates,tibshirani,ovadia,odin}.

This concern is important for jailbreak detection because curated instruction datasets can differ substantially from real-user chatbot traffic \cite{toxicchat,lmsyschat,wildchat}. As a result, a detector that appears well calibrated on curated benign prompts may operate at a very different FPR in deployment-like settings. We refer to the ability of a threshold to preserve its intended false-positive budget across benign distributions as \emph{calibration transfer}. This motivates our evaluation of FPR on real-user distributions and the resulting family-level detection at low-FPR operating points. Section~\ref{sec:method} formalizes this calibration dependence together with reference sensitivity.

\section{Data-Dependent Evaluation Framework}
\label{sec:method}

We evaluate model-internal jailbreak detectors as data-dependent decision systems. In addition to the query prompt, their behavior can depend on two external data sources: reference data that influence the detector score and benign calibration data that determine the operating threshold. Our methodology separates these dependencies and measures their effects on ranking quality and attack-family coverage.

\subsection{Detector Decision Model}

Let $x$ denote a query prompt, $R$ the detector's reference data, and $s(x;R)$ its continuous detection score. The form of $R$ depends on the detector. For example, MTK uses a bank of benign and malicious anchors, GradSafe uses safe and unsafe gradient references, and a supervised probe learns its parameters from labeled reference representations. For detectors without explicit reference data, $R$ can be treated as fixed or empty.

Given a benign calibration set $C$ and a target false-positive budget $\alpha$, we define the operating threshold as
\begin{equation}
\tau_{\alpha}(R,C)
=
Q_{1-\alpha}
\left(
\{s(x;R):x\in C\}
\right),
\label{eq:threshold}
\end{equation}
where $Q_{1-\alpha}$ denotes the empirical $(1-\alpha)$ quantile. The detector decision is
\begin{equation}
D_{\alpha}(x;R,C)
=
\mathbb{I}
\left[
s(x;R)>\tau_{\alpha}(R,C)
\right].
\label{eq:decision}
\end{equation}

Changing $C$ changes the operating threshold for a fixed scoring function, whereas changing $R$ can change both query scores and the benign score distribution used to estimate the threshold. We use $\alpha=0.05$ for comparison with prior detector evaluations and $\alpha=0.01$ as the stricter deployment-oriented operating point.

\subsection{Calibration Transfer}

We study whether a threshold calibrated on one benign distribution preserves its intended false-positive budget on another. For a disjoint benign evaluation set $B$, the realized FPR is \begin{equation}
\widehat{\mathrm{FPR}}_{B}(R,C)
=
\frac{1}{|B|}
\sum_{x\in B}
D_{\alpha}(x;R,C).
\end{equation}
We characterize calibration transfer by comparing the realized rate with the nominal budget $\alpha$, while holding the detector and reference configuration fixed.

\subsection{Reference-Set Sensitivity}

We study how reference composition affects detector behavior. For an attack family $A$, its TPR under reference set $R$ and calibration set $C$ is
\begin{equation}
\mathrm{TPR}_{A}(R,C)
=
\frac{1}{|A|}
\sum_{x\in A}
D_{\alpha}(x;R,C).
\end{equation}
We compare $\mathrm{TPR}_{A}(R,C)$ across reference draws and causal reference interventions, recalibrating the threshold for each reference configuration. Random redraws measure sensitivity to reference selection, whereas controlled interventions isolate the effect of deliberately changing reference composition while holding the remaining evaluation conditions
fixed.

We distinguish two effects. Changes in AUROC measure changes in attack--benign ranking, whereas changes in $\mathrm{TPR}_{A}(R,C)$ measure changes in family-level coverage at the selected operating point. This distinction separates ranking sensitivity from operating-point sensitivity.

\subsection{MTK as a Mechanistic Case Study}

MTK makes reference dependence explicit through its layer-wise rank feature. For prompt $p$ at layer $\ell$, let $m_j$ denote the number of malicious anchors ranked ahead of the $j$-th nearest benign anchor. MTK computes \begin{equation}
r_\ell(p)
=
\frac{k+1}{2}
+
\frac{1}{k}
\sum_{j=1}^{k}m_j.
\label{eq:identity}
\end{equation}
The feature therefore depends on the interleaving of malicious anchors among the nearest benign references rather than on absolute distance alone. Changing the reference bank can alter these interleavings, the layer-wise rank trajectory, and ultimately the detector score. We use reference-bank redraws to measure sensitivity and targeted anchor interventions to test the causal effect of reference composition on attack-family coverage.

\section{Experimental Setup}
\label{sec:setup}

This section specifies the models, detector configurations, datasets, calibration settings, and experimental controls used to instantiate our framework. The code and evaluation artifacts are publicly available. \footnote{\url{https://github.com/Mtehabsim/internal-detector-sensititvity}}

\subsection{Models and Detector Configurations}

We evaluate the four chat models supported by the MTK release: \mbox{Llama-2}-7B-chat \cite{llama2}, \mbox{Llama-3}-8B-Instruct \cite{llama3}, Mistral-7B-Instruct-v0.2 \cite{mistral}, and Vicuna-7B-v1.5 \cite{vicuna}. We evaluate the latest MTK release (commit \texttt{c5e2f18}, September 2026) using the model-specific configured seeds 27, 143, 47, and 56, respectively.

Our evaluation invokes MTK's released representation-extraction, ranking, and Isolation Forest \cite{iforest} components without modifying its scoring procedure. At the configured seeds, it reproduces the release's per-prompt scores and per-family AUROCs. We preserve the model-specific representation layers used in the release, including the released Mistral configuration in which query and anchor representations are offset by one transformer block.

MTK uses a reference bank of 1,600 prompts, equally divided between benign and malicious anchors. The benign anchors comprise 300 Dolly \cite{dolly}, 300 Alpaca \cite{alpaca}, and 200 OR-Bench \cite{orbench} prompts; the malicious anchors comprise 100 AdvBench \cite{advbench}, 100 MaliciousInstruct \cite{maliciousinstruct}, and 600 PKU-SafeRLHF \cite{pku} prompts.

We additionally evaluate GradSafe, HiddenDetect, a supervised linear probe, and windowed perplexity. For \mbox{Llama-2}, we use the official zero-shot GradSafe configuration \cite{gradsafe}; for the remaining models, we apply the same procedure using their corresponding chat templates. The \mbox{Llama-2} implementation reproduces the official reference gradients and selected parameter dimensions. HiddenDetect \cite{hiddendetect} is evaluated using the LLM adaptation over layers 16--29 employed in the MTK baseline evaluation.

The linear probe is an $L_2$-regularized logistic-regression classifier trained on the 800 benign and 800 malicious MTK anchors, with $C=1$. We select its representation layer through five-fold stratified cross-validation on the reference data and refit the probe on all anchors at the selected layer. Evaluation attacks are excluded from model selection. Windowed perplexity \cite{alon,jain} serves as a surface-level baseline that operates directly on the input sequence.

\subsection{Attack and Benign Datasets}

\textbf{Attack prompts.}
We evaluate the twelve jailbreak families included in the MTK release. They span suffix-based attacks (SAA \cite{saa} and nanoGCG \cite{advbench}), multilingual transformation (Zulu \cite{zulu}), evolved jailbreak templates (AutoDAN \cite{autodan}), request decomposition (DrAttack \cite{drattack}), in-the-wild jailbreaks (IJP \cite{ijp}), attacker-LLM refinement (PAIR \cite{pair} and TAP \cite{tap}), persuasive paraphrasing (three generator-specific PAP variants \cite{pap}), and the JailJudge prompt family \cite{jailjudge}.

The released SAA data exhibit limited request diversity because their prompt texts share the same underlying request and vary primarily in their adversarial suffixes. We therefore construct \emph{\mbox{SAA-850}} to evaluate whether detector behavior extends across distinct harmful requests. Each example combines a different request, one per record of the release's SAA file (520 AdvBench behaviors \cite{advbench} and 330 HEx-PHI instructions \cite{hexphi}), with the corresponding released SAA template and that record's own suffix, producing 850 distinct requests under a common attack construction. We also evaluate the same requests without the suffix. The released suffixes are transferred without re-optimization.

\textbf{Benign and calibration prompts.}
We use Dolly, ToxicChat, and WildChat to represent curated and deployment-like benign distributions. OR-Bench provides an additional difficult-benign control set. Table~\ref{tab:benign_data} summarizes their experimental roles.

\begin{table}[t]
\centering
\small
\caption{Benign datasets used for calibration and evaluation. Calibration and evaluation partitions are disjoint.}
\label{tab:benign_data}
\setlength{\tabcolsep}{2pt}
\renewcommand{\arraystretch}{1}
\begin{tabularx}{\columnwidth}{@{}l X c c@{}}
\toprule
\textbf{Dataset} & \textbf{Type} & \textbf{Calibration} & \textbf{Evaluation} \\
\midrule
Dolly \cite{dolly}
    & Curated instructions
    & 200
    & Held-out \\
ToxicChat \cite{toxicchat}
    & Real-user prompts
    & $\sim$1K
    & 500 \\
WildChat \cite{wildchat}
    & Real-user prompts
    & 1K
    & 500 \\
OR-Bench \cite{orbench}
    & Difficult benign
    & --
    & Control \\
\bottomrule
\end{tabularx}
\end{table}

For Dolly, we exclude prompts already present in MTK's benign reference bank. ToxicChat serves as the primary deployment-like distribution, while WildChat provides an independent real-user distribution. All partitions are deduplicated, and overlaps between calibration and evaluation sets are removed.

\subsection{Calibration and Evaluation Protocol}

We evaluate both 1\% and 5\% false-positive budgets. The 5\% operating point supports comparison with MTK and prior detector evaluations, while 1\% provides the stricter deployment-oriented setting emphasized in our cross-detector analysis. We consider two calibration settings. The \emph{curated} setting calibrates the threshold on 200 Dolly prompts and evaluates it on held-out Dolly prompts. The \emph{deployment-like} setting calibrates the threshold on real-user prompts and evaluates the realized FPR on a disjoint real-user set.

MTK's main evaluation also uses ToxicChat benign prompts and describes calibration on a portion of its validation set. Because the released LLM detector does not provide the corresponding thresholding procedure, we cannot reconstruct that setting exactly. We therefore use disjoint ToxicChat calibration and evaluation partitions as the closest available approximation.

Following Section~\ref{sec:method}, we report the realized FPR, per-family detection rate, and AUROC at each evaluated operating point, with uncertainty estimated using 2{,}000 bootstrap resamples. HarmBench \cite{harmbench} and StrongREJECT \cite{strongreject} analyses are provided in the artifact because our manual audit found that their labels did not reliably track actionable harm. We therefore interpret TPR as detection of attack prompts rather than harmful outputs prevented.
\begin{table*}[t]
\centering
\caption{Calibration transfer from curated to real-user benign prompts.
Thresholds are calibrated to a nominal 5\% false-positive budget using either
Dolly or ToxicChat (TC). FPR$_{\mathrm{TC}}$ and FPR$_{\mathrm{WC}}$ denote
the realized false-positive rates on held-out ToxicChat and WildChat,
respectively; Det. is the mean detection rate across the twelve attack
families.}

\label{tab:calib}
\footnotesize
\setlength{\tabcolsep}{3pt}
\begin{tabular}{lrrrrrrrrrrrrrrrr}
\toprule
 & \multicolumn{4}{c}{Llama-2} & \multicolumn{4}{c}{Llama-3} & \multicolumn{4}{c}{Mistral} & \multicolumn{4}{c}{Vicuna} \\
\cmidrule(lr){2-5} \cmidrule(lr){6-9} \cmidrule(lr){10-13} \cmidrule(lr){14-17}
Threshold from & Dolly & Dolly & Dolly & TC & Dolly & Dolly & Dolly & TC & Dolly & Dolly & Dolly & TC & Dolly & Dolly & Dolly & TC \\
Measured & FPR$_\mathrm{TC}$ & FPR$_\mathrm{WC}$ & Det. & Det. & FPR$_\mathrm{TC}$ & FPR$_\mathrm{WC}$ & Det. & Det. & FPR$_\mathrm{TC}$ & FPR$_\mathrm{WC}$ & Det. & Det. & FPR$_\mathrm{TC}$ & FPR$_\mathrm{WC}$ & Det. & Det. \\
\midrule
MTK & 0.236 & 0.286 & 0.965 & 0.570 & 0.116 & 0.136 & 0.933 & 0.742 & 0.272 & 0.368 & 0.968 & 0.608 & 0.252 & 0.284 & 0.909 & 0.554 \\
Linear probe & 0.198 & 0.232 & 0.954 & 0.705 & 0.124 & 0.174 & 0.880 & 0.683 & 0.166 & 0.256 & 0.765 & 0.506 & 0.164 & 0.152 & 0.765 & 0.403 \\
GradSafe & 0.078 & 0.096 & 0.775 & 0.697 & 0.110 & 0.186 & 0.856 & 0.646 & 0.304 & 0.286 & 0.754 & 0.350 & 0.260 & 0.258 & 0.744 & 0.410 \\
HiddenDetect & 0.078 & 0.100 & 0.620 & 0.488 & 0.068 & 0.052 & 0.131 & 0.085 & 0.026 & 0.016 & 0.037 & 0.094 & 0.098 & 0.120 & 0.772 & 0.529 \\
Windowed PPL & 0.162 & 0.156 & 0.343 & 0.252 & 0.122 & 0.164 & 0.286 & 0.233 & 0.138 & 0.166 & 0.357 & 0.213 & 0.174 & 0.164 & 0.226 & 0.121 \\
\bottomrule
\end{tabular}
\end{table*}

\subsection{Reference-Set Controls}

The MTK random seed jointly determines the reference bank, sampled attack prompts, and the isolation forest. Consequently, varying the seed alone does not isolate reference-set effects. We therefore separate observational sensitivity from causal intervention. First, we repeat the released evaluation over seeds 0--49 and include the configured model seed when it lies outside this range. These runs characterize variability across complete released configurations. Second, we hold the evaluation prompts and other randomized detector components fixed
while redrawing only the reference bank. These bank-only redraws isolate sensitivity to reference composition but remain observational.

To test the causal effect of reference composition, we perform targeted interventions in which selected malicious anchors are replaced with SAA-shaped or control anchors while preserving bank size, replacement positions, and disjointness between reference and evaluation prompts. Each intervention refits the Isolation Forest and recalibrates the operating
threshold. For GradSafe, we similarly evaluate multiple safe and unsafe reference sets while holding the remaining evaluation procedure fixed.

\subsection{Threat Model}

We consider a non-adaptive attacker who submits published jailbreak prompts without access to or explicit optimization against the detector. The defender operates the detector under a specified benign false-positive budget. This setting tests whether known jailbreak families receive consistent coverage before the
attacker adapts to the defense.

Adaptive monitor-evasion attacks are outside the scope of this study. Prior work shows that detector-aware adversaries can explicitly optimize internal representations to evade latent-space monitors \cite{bailey,nasr}. Such attacks constitute a stronger threat model; our evaluation asks whether deployment limitations arise even
under public, non-adaptive attacks.

\section{Evaluation and Results}
\label{sec:res}

We organize the evaluation around three questions: (1) whether calibration transfers
from curated to deployment-like benign traffic, (2) whether reference composition
causally affects detector behavior, and (3) whether the resulting coverage gaps
persist across detector designs. We use the 5\% false-positive budget for
comparison with the released detector evaluations and for the reference and mechanistic analyses. We additionally evaluate a stricter 1\% budget as a more
deployment-oriented operating point, where false alarms impose greater
operational and analyst costs.


\subsection{Calibration Transfer}
\label{sec:r1}

We first examine whether a threshold calibrated on curated benign prompts
preserves its intended false-positive budget on real-user traffic.
As a consistency check, the Dolly-calibrated MTK configuration approximately
reproduces the reported operating behavior on curated prompts. On
\mbox{Llama-2}, the threshold detects \nCalibLtwoNanoDolly\ of the evaluated
nanoGCG prompts, compared with 0.95 reported by MTK, while the
false-positive rate on held-out Dolly prompts remains close to the nominal
5\% budget. Because the released implementation does not expose every detail
of the original thresholding procedure, we treat this as a consistency check
rather than an exact reproduction.

\subsubsection{Curated Calibration Does Not Transfer to Real-User Traffic}

The same thresholds do not preserve the 5\% budget on real-user prompts. For
MTK, the realized FPR on ToxicChat ranges from
\nCalibMTKFPRrealMin\ to \nCalibMTKFPRrealMax\ across the four models,
corresponding to \nCalibMTKFPRratioMin--\nCalibMTKFPRratioMax\ times the
nominal budget. The effect extends beyond MTK, as the supervised probe, GradSafe,
and windowed perplexity exceed the target on ToxicChat in all four models,
while HiddenDetect exceeds it in three of four models as shown in Table~\ref{tab:calib}. Because the threshold is estimated from only 200 calibration prompts, some deviation from the nominal 5\% rate is expected by chance. We therefore compare the realized FPRs with a one-sided 95\% exchangeability bound of approximately 9\%; most detector--model cases remain above this bound even after correcting for multiple comparisons.

\begin{figure}[t]
\centering
\includegraphics[width=\columnwidth]{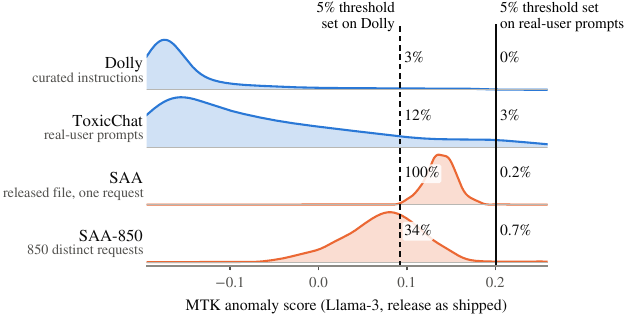}
\caption{Calibration transfer and attack coverage under curated and real-user thresholds.}
\label{fig:calibration}
\end{figure}

For MTK and the supervised probe, Dolly also contributes
to the benign reference data. As shown in Fig.~\ref{fig:calibration},
Dolly prompts receive lower scores than the real-user prompts,
yielding a lower operating threshold and thereby increasing
the realized FPR after transfer.

\begin{table*}[t]
\centering
\caption{MTK per-family ranking and detection under real-user calibration.
TPR is measured at the nominal 5\% false-positive budget. Bold values indicate
TPR $<0.20$ when AUROC $\geq0.85$.}
\label{tab:r1}

\footnotesize
\setlength{\tabcolsep}{4.5pt}
\renewcommand{\arraystretch}{1.08}

\begin{tabular}{@{}l c cc cc cc cc@{}}
\toprule
& &
\multicolumn{2}{c}{\textbf{Llama-2}} &
\multicolumn{2}{c}{\textbf{Llama-3}} &
\multicolumn{2}{c}{\textbf{Mistral}} &
\multicolumn{2}{c}{\textbf{Vicuna}} \\
\cmidrule(lr){3-4}
\cmidrule(lr){5-6}
\cmidrule(lr){7-8}
\cmidrule(lr){9-10}

\textbf{Attack Family} &
\textbf{$n$} &
\textbf{AUROC} & \textbf{TPR@5\%} &
\textbf{AUROC} & \textbf{TPR@5\%} &
\textbf{AUROC} & \textbf{TPR@5\%} &
\textbf{AUROC} & \textbf{TPR@5\%} \\
\midrule

SAA
& 500
& 0.942 & \textbf{0.114}
& 0.918 & \textbf{0.002}
& 0.981 & 1.000
& 0.850 & \textbf{0.026} \\

Zulu
& 500
& 0.902 & \textbf{0.034}
& 0.975 & 0.752
& 0.958 & 0.688
& 0.866 & \textbf{0.108} \\

AutoDAN
& 500
& 0.969 & 0.936
& 0.984 & 0.956
& 0.925 & \textbf{0.108}
& 0.961 & 0.744 \\

nanoGCG
& 42/500/500/500
& 0.914 & 0.429
& 0.991 & 0.952
& 0.978 & 0.876
& 0.985 & 0.922 \\

DrAttack
& 500
& 0.922 & 0.584
& 0.960 & 0.862
& 0.928 & \textbf{0.124}
& 0.957 & 0.788 \\

IJP
& 80
& 0.981 & 0.963
& 0.982 & 0.938
& 0.985 & 0.938
& 0.936 & 0.725 \\

JailJudge
& 500
& 0.895 & 0.548
& 0.976 & 0.876
& 0.918 & 0.550
& 0.923 & 0.512 \\

PAIR
& 500
& 0.936 & 0.768
& 0.956 & 0.828
& 0.977 & 0.906
& 0.939 & 0.726 \\

TAP
& 50
& 0.944 & 0.800
& 0.975 & 0.920
& 0.955 & 0.820
& 0.946 & 0.820 \\

PAP (GPT-3.5)
& 19
& 0.949 & 0.737
& 0.953 & 0.684
& 0.913 & 0.368
& 0.866 & 0.526 \\

PAP (GPT-4)
& 44
& 0.951 & 0.682
& 0.958 & 0.727
& 0.940 & 0.591
& 0.867 & 0.477 \\

PAP (Llama-2)
& 37
& 0.916 & 0.243
& 0.922 & 0.405
& 0.860 & 0.324
& 0.845 & 0.270 \\

\midrule
\multicolumn{10}{@{}l}{\textbf{SAA Generalization and Suffix Ablation}} \\

SAA-850 (transferred suffix)
& 850
& 0.940 & \textbf{0.112}
& 0.851 & \textbf{0.007}
& 0.978 & 0.999
& 0.745 & 0.018 \\

SAA-850 (suffix removed)
& 850
& 0.979 & 0.935
& 0.986 & 0.969
& 0.978 & 0.995
& 0.781 & 0.067 \\

\midrule
\multicolumn{10}{@{}l}{\textbf{Aggregate and Operating-Point Statistics}} \\

Mean (12 families)
& --
& 0.935 & 0.570
& 0.962 & 0.742
& 0.943 & 0.608
& 0.912 & 0.554 \\

Realized FPR
& 500
& \multicolumn{2}{c}{0.054}
& \multicolumn{2}{c}{0.032}
& \multicolumn{2}{c}{0.048}
& \multicolumn{2}{c}{0.042} \\

\bottomrule
\end{tabular}

\vspace{2pt}
\begin{minipage}{0.98\textwidth}
\scriptsize
\textit{Note:} nanoGCG uses 42/500/500/500 prompts for
Llama-2/Llama-3/Mistral/Vicuna, respectively.
\end{minipage}

\end{table*}

The same pattern appears on WildChat. Dolly-calibrated thresholds exceed the
target in \nWCDollyAbove\ of \nWCPairs\ detector--model combinations.
Removing WildChat prompts containing explicit jailbreak markers produces
little change, indicating that the mismatch is not driven by a small number of
obviously suspicious prompts. Thresholds calibrated on ToxicChat and WildChat
transfer more reliably between the two real-user distributions than thresholds
calibrated on Dolly, although neither transfer is exact.


This difference is not explained by calibration-set size. When MTK is
calibrated on ToxicChat subsets matched to the Dolly calibration-set size, the
resulting FPR remains below that produced by the Dolly-calibrated threshold.
The transfer gap therefore reflects the calibration distribution rather than
the number of calibration examples.

\subsubsection{Real-User Recalibration Exposes Family-Level Gaps}

Recalibrating on real-user prompts brings the realized FPR closer to the
nominal 5\% budget, but does not preserve attack coverage uniformly.
Table~\ref{tab:r1} shows that several attack families retain high AUROC while
their TPR at the calibrated operating point is low.

\begin{figure}[t]
\centering
\includegraphics[width=\columnwidth]{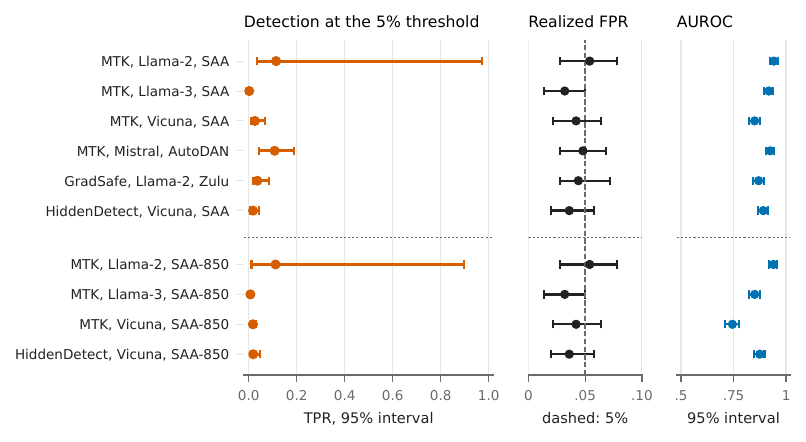}
\caption{Ranking quality and operating-point detection at the 5\%
real-user-calibrated threshold. Error bars show bootstrap uncertainty.}
\label{fig:calibrated-results}
\end{figure}

SAA provides a clear example. Its AUROC is \nLtwoSAAAUROC\ on
\mbox{Llama-2} and \nLthreeSAAAUROC\ on \mbox{Llama-3}, yet its TPR is only
\nLtwoSAATPR\ and \nLthreeSAATPR, respectively. Mistral detects SAA at its configured seed (median \nMisLotterySAATPRmed\ across the 50 release seeds), but exhibits low coverage for other families, including AutoDAN and DrAttack.

The \mbox{SAA-850} transfer set shows that this behavior is not limited to the
single underlying request in the released SAA data. At the 5\% operating
point, MTK detects \nLtwoSAAeightTPR, \nLthreeSAAeightTPR, and
\nVicSAAeightTPR\ of \mbox{SAA-850} on \mbox{Llama-2}, \mbox{Llama-3},
and Vicuna, respectively. Removing the SAA suffix substantially increases
detection on the Llama models, showing that the coverage gap persists across
diverse harmful requests under the transferred SAA construction.


Figure~\ref{fig:calibrated-results} emphasizes that high AUROC does not imply
high TPR at the selected operating point. It also distinguishes stable misses,
such as \mbox{Llama-3} SAA, from cases such as \mbox{Llama-2} SAA whose TPR is
more sensitive to finite-sample threshold variation.

\begin{figure*}[t]
\centering
\includegraphics[width=\textwidth]{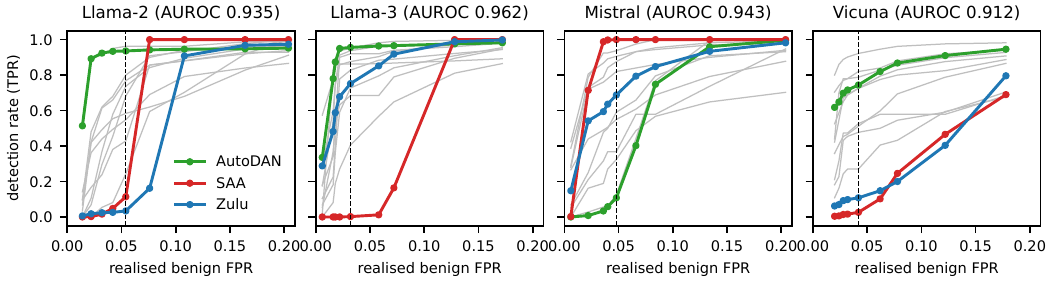}
\caption{MTK family-level detection versus realized FPR under real-user calibration. Gray curves denote the remaining attack families; the dashed line marks the nominal 5\% operating point.}
\label{fig:budget}
\end{figure*}

Figure~\ref{fig:budget} shows why this distinction matters. Several families
exhibit sharp changes in TPR over a narrow range of false-positive rates. For
\mbox{Llama-2} SAA, detection increases from \nLtwoSAATPR\ to
\nLtwoSAATPRsevenfive\ as the realized FPR changes from \nLtwoFPR\ to
\nLtwoFPRsevenfive. Thus, a detector can rank a family well globally while
providing little coverage at the low-FPR threshold used in deployment.
As an interpretive check, an equal-variance Gaussian reference with the same
AUROC would yield substantially higher detection for several of these
high-AUROC/low-TPR cases. The empirical gap therefore reflects not only
ranking quality, but also the placement and concentration of family scores
relative to the operating threshold.

\subsection{Reference-Set Sensitivity}
\label{sec:r2}

We next examine whether the reference data themselves affect family-level
coverage. Random redraws provide evidence of sensitivity, while causal
reference interventions test whether reference composition directly changes detector behavior.

\subsubsection{Variation Across Reference Draws}

Across release-seed runs, aggregate AUROC varies relatively little, yet
family-level detection can change substantially. Every evaluated draw contains
at least one attack family with 50 or more prompts whose TPR falls below 0.2
at the 5\% operating point. Mistral provides the clearest example: SAA
detection ranges from \nMisLotterySAATPRmin\ to \nMisLotterySAATPRmax\ across
draws even though aggregate AUROC remains relatively stable.
The released seed changes several components simultaneously, including the
reference bank, evaluation sample, and Isolation Forest. We therefore repeat
the experiment while fixing the evaluation prompts and forest and redrawing
only the reference bank. Mistral SAA detection still ranges from
\nMisBankOnlySAATPRmin\ to \nMisBankOnlySAATPRmax. Thus, changing reference
composition alone is sufficient to change detection on fixed prompts.

The observed bank-to-bank variation does not consistently exceed prompt-sampling uncertainty. We therefore interpret these redraws as evidence of reference sensitivity, but not as evidence that reference selection is always the dominant source of variance.

\begin{figure*}[t]
\centering
\includegraphics[scale=0.9]{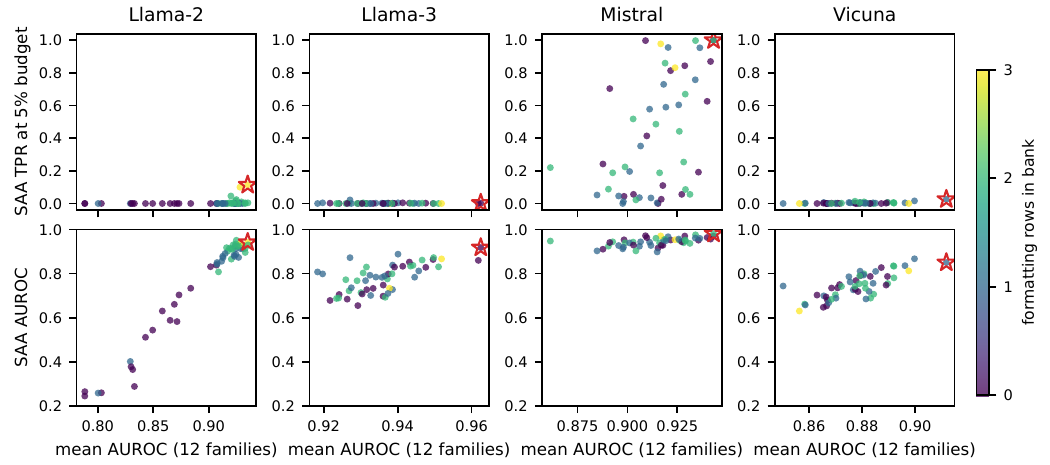}
\caption{MTK sensitivity across release seeds. Each point denotes one run and
stars mark the configured seeds. The top panel relates aggregate AUROC to SAA
detection at the 5\% operating point; the bottom panel relates SAA AUROC to
formatting artifacts sampled into the malicious reference bank.}
\label{fig:lottery}
\end{figure*}

Figure~\ref{fig:lottery} also shows that the configured seeds lie near the upper end of the observed aggregate-AUROC distributions. By mean AUROC, the configured seed ranks second among 50 runs for \mbox{Llama-2} and first for \mbox{Llama-3}, Mistral, and Vicuna. We do not infer why these seeds were selected; the result instead motivates reporting performance across multiple reference draws rather than relying on a single configured instance.

\subsubsection{Causal Reference Interventions}

Random reference redraws establish an association between reference composition
and detector behavior, but they do not show whether a specific change in the
reference bank causes the observed coverage shift. We therefore perform
controlled interventions in which 200 of the 600 PKU-SafeRLHF \cite{pku} malicious
anchors are replaced while preserving the bank size and replacement positions.
As a control, we replace these anchors with ordinary PKU-SafeRLHF prompts; in
the targeted intervention, we replace them with SAA-shaped anchors constructed
from prompts disjoint from the SAA evaluation set. Each intervention refits the
Isolation Forest and recalibrates the operating threshold.
\begin{table}[t]
\centering
\caption{Controlled reference interventions and family-level coverage at the
recalibrated 5\% operating point. Each entry reports SAA TPR / mean TPR over
the remaining eleven attack families. Each entry is a mean of three banks; Ordinary PKU is the control.}
\label{tab:sentinel}

\scriptsize
\setlength{\tabcolsep}{10pt}
\renewcommand{\arraystretch}{1.45}

\begin{tabular}{@{}lcccc@{}}
\toprule
\textbf{Intervention}
& \textbf{Llama-2}
& \textbf{Llama-3}
& \textbf{Mistral}
& \textbf{Vicuna} \\
\midrule

None
& .040/.610
& .002/.785
& .438/.472
& .011/.563 \\

Ordinary PKU
& .000/.555
& .002/.789
& .357/.417
& .009/.559 \\

SAA-shaped
& \textbf{.706}/.564
& \textbf{1.000}/.727
& .324/.095
& \textbf{1.000}/.492 \\

SAA + shuffle
& .670/.570
& \textbf{1.000}/.728
& .339/.120
& \textbf{1.000}/.447 \\

SAA + random
& .376/.586
& \textbf{1.000}/.762
& .000/.138
& \textbf{1.000}/.421 \\

\bottomrule
\end{tabular}
\end{table}

Table~\ref{tab:sentinel} shows that targeted changes in reference composition causally alter family-level coverage. On Llama-3, SAA TPR remains 0.002 after the ordinary
PKU replacement but increases from 0.002 to 1.000 when SAA-shaped references
are introduced. Vicuna shows the same contrast: SAA TPR changes only from
0.011 to 0.009 under the ordinary replacement, but rises to 1.000 under each
SAA-shaped intervention. Llama-2 also shows a substantial increase, from
0.040 at baseline to 0.706 with the SAA-shaped intervention, with the shuffled
and random variants yielding TPRs of 0.670 and 0.376, respectively.

The intervention is not uniformly beneficial across models or attack families.
On Mistral, baseline SAA TPR is 0.438, but it decreases to 0.324, 0.339, and
0.000 under the three SAA-shaped variants. Coverage of the remaining eleven
families also declines substantially, with mean TPR falling from 0.472 at
baseline to 0.095--0.138 across these interventions. Smaller
trade-offs also appear on Llama-3 and Vicuna: increasing SAA TPR to 1.000 reduces
mean TPR on the other families from 0.785 to 0.727--0.762 on Llama-3 and from
0.563 to 0.421--0.492 on Vicuna.
These controlled interventions show that the reference bank does more than
provide a fixed comparison set. Its composition causally changes family-level coverage at the calibrated operating point, and improving
coverage of a targeted family can redistribute or reduce coverage elsewhere.

\subsubsection{Mechanistic Evidence from MTK}

MTK allows us to trace this reference dependence directly because its rank
feature is defined by the ordering of benign and malicious anchors around each
query. Reconstructing the feature from Eq.~\eqref{eq:identity} reproduces the
released values to within $\nIdentityMaxDiff$.

For \mbox{Llama-2}, \mbox{Llama-3}, and Vicuna, SAA prompts exhibit fewer
malicious anchors ahead of their nearest benign neighbors than benign prompts
do; Mistral shows the opposite pattern, consistent with its stronger SAA
coverage. This indicates that MTK's Isolation Forest detects deviations from
benign rank trajectories rather than a monotonic notion of maliciousness.

We also identify a concrete reference-data quality issue. Three formatting artifacts from the PKU-SafeRLHF malicious pool occur in the configured \mbox{Llama-2} bank and frequently rank near SAA prompts. Across reference draws, SAA AUROC increases with the number of these artifacts (Spearman's $\rho=\nLtwoMarkupSpearman$, $n=50$). Replacing them with ordinary malicious anchors reduces SAA AUROC, whereas replacing ordinary anchors has a much smaller effect. The artifacts primarily affect ranking rather than the 5\% TPR failure, showing that small reference-data irregularities can measurably alter MTK's rank geometry.

\subsection{Low-FPR Behavior Across Detectors}
\label{sec:r3}

The preceding experiments establish calibration and reference effects
primarily through MTK. We finally examine whether incomplete family coverage persists across detector designs.

\begin{table*}[t]
\centering
\caption{Cross-detector performance at 1\% and 5\% false-positive budgets.
Results report mean detection across the twelve attack families, SAA-specific
performance, realized false-positive rates. AUROC is reported separately to distinguish ranking
quality from operating-point detection. Not a ranking of detectors: GradSafe outside Llama-2 is our port, and HiddenDetect is our adaptation.}
\label{tab:budget}
\footnotesize
\setlength{\tabcolsep}{10pt}
\begin{tabular}{llrrrrrrrr}
\toprule
 & & \multicolumn{3}{c}{Mean over 12 families} & \multicolumn{3}{c}{SAA} & \multicolumn{2}{c}{FPR} \\
\cmidrule(lr){3-5} \cmidrule(lr){6-8} \cmidrule(lr){9-10}
Model & Detector & AUROC & 1\% & 5\% & AUROC & 1\% & 5\% & 1\% & 5\% \\
\midrule
Llama-2 & MTK & 0.935 & 0.125 & 0.570 & 0.942 & 0.000 & 0.114 & 0.014 & 0.054 \\
 & GradSafe & 0.937 & 0.220 & 0.697 & 0.957 & 0.000 & 0.812 & 0.010 & 0.044 \\
 & HiddenDetect & 0.893 & 0.079 & 0.488 & 0.765 & 0.000 & 0.000 & 0.012 & 0.044 \\
 & Linear probe & 0.950 & 0.363 & 0.705 & 0.882 & 0.000 & 0.002 & 0.012 & 0.040 \\
 & Windowed PPL & 0.574 & 0.171 & 0.252 & 1.000 & 1.000 & 1.000 & 0.006 & 0.060 \\
\midrule
Llama-3 & MTK & 0.962 & 0.174 & 0.742 & 0.918 & 0.000 & 0.002 & 0.006 & 0.032 \\
 & GradSafe (port) & 0.947 & 0.200 & 0.646 & 0.832 & 0.000 & 0.008 & 0.006 & 0.018 \\
 & HiddenDetect & 0.591 & 0.012 & 0.085 & 0.090 & 0.000 & 0.000 & 0.008 & 0.040 \\
 & Linear probe & 0.938 & 0.416 & 0.683 & 0.586 & 0.000 & 0.000 & 0.004 & 0.030 \\
 & Windowed PPL & 0.595 & 0.160 & 0.233 & 1.000 & 1.000 & 1.000 & 0.010 & 0.048 \\
\midrule
Mistral & MTK & 0.943 & 0.160 & 0.608 & 0.981 & 0.002 & 1.000 & 0.006 & 0.048 \\
 & GradSafe (port) & 0.786 & 0.178 & 0.350 & 0.849 & 0.000 & 0.012 & 0.008 & 0.050 \\
 & HiddenDetect & 0.626 & 0.008 & 0.094 & 0.438 & 0.000 & 0.000 & 0.006 & 0.072 \\
 & Linear probe & 0.881 & 0.286 & 0.506 & 0.968 & 0.000 & 0.466 & 0.008 & 0.028 \\
 & Windowed PPL & 0.588 & 0.123 & 0.213 & 1.000 & 1.000 & 1.000 & 0.004 & 0.052 \\
\midrule
Vicuna & MTK & 0.912 & 0.369 & 0.554 & 0.850 & 0.004 & 0.026 & 0.020 & 0.042 \\
 & GradSafe (port) & 0.806 & 0.213 & 0.410 & 0.930 & 0.098 & 0.298 & 0.018 & 0.064 \\
 & HiddenDetect & 0.930 & 0.268 & 0.529 & 0.892 & 0.000 & 0.018 & 0.006 & 0.036 \\
 & Linear probe & 0.889 & 0.188 & 0.403 & 0.885 & 0.000 & 0.000 & 0.012 & 0.058 \\
 & Windowed PPL & 0.521 & 0.085 & 0.121 & 0.726 & 0.252 & 0.252 & 0.016 & 0.050 \\
\bottomrule
\end{tabular}
\end{table*}

\subsubsection{Family-Level Coverage}

Table~\ref{tab:budget} reports performance at both 1\% and 5\% false-positive
budgets. We retain 5\% for comparison with prior detector evaluations,
while emphasizing 1\% as the stricter deployment-oriented operating point.

At the nominal 5\% target, 11 of the 16 detector--model pairs contain at least one family of 50 or more prompts with AUROC $\geq 0.85$ and TPR $< 0.20$. Excluding SAA, \nBudgetModelsWithHiddenFamilyNoSAA\ pairs still contain such a family, with Zulu accounting for \nBudgetNoSAAHiddenZulu\ of them.

The stricter 1\% operating point exposes a larger practical gap. No evaluated
internal detector exceeds a mean family-level TPR of \nBudgetTopMeanTPRone\,
despite mean AUROC reaching \nBudgetMaxMeanAUROC. Thus, the ranking performance
suggested by AUROC does not translate directly into useful attack coverage
when false alarms must be tightly constrained.

\subsubsection{Ranking Failures Versus Operating-Point Failures}

Low TPR arises for different reasons across detector designs. HiddenDetect on
several models and the \mbox{Llama-3} supervised probe rank SAA poorly,
indicating a ranking failure. In contrast, MTK on \mbox{Llama-2} and
\mbox{Llama-3} achieves SAA AUROC above 0.90 but detects little of the family
at the 5\% operating point. These are operating-point failures because the ordering is
strong overall, but the attack scores remain below the threshold required to
control benign false positives.

Failure patterns also differ across detectors. MTK detects SAA at its configured seed on Mistral, while official GradSafe provides substantially higher SAA coverage on Llama-2. The supervised probe achieves strong cross-validation AUROC on
its reference anchors but still leaves substantial gaps on attack families
absent from its training data. These differences indicate that no single
internal representation or detector design uniformly covers the evaluated
jailbreak families.

Windowed perplexity provides complementary coverage for several
high-perplexity suffix attacks but performs less effectively on fluent attack
families. We therefore treat it as a useful complementary signal rather than a
replacement for low-FPR evaluation of model-internal detectors.

\section{Discussion}
\label{sec:disc}
Our results show that model-internal jailbreak detectors should be evaluated at
explicit low-FPR operating points rather than relying on aggregate AUROC alone.
Evaluations should report the calibration distribution, realized FPR on
deployment-like traffic, and per-family TPR at stated false-positive budgets.
For reference-based detectors, performance should also be evaluated across
multiple reference draws because reference composition can change family-level
coverage. Our causal interventions further show that improving coverage of
one attack family can reduce coverage of others. These findings motivate
reporting TPR at strict operating points, particularly TPR@1\% FPR for
deployment-oriented evaluation, while retaining 5\% operating point for comparison with
prior work.

Several limitations remain. ToxicChat and WildChat provide useful proxies for
real-user traffic but do not represent all deployment environments, and
WildChat benign labels are automatically derived. Bootstrap intervals capture
finite-sample uncertainty within the observed attack families but do not
measure generalization to unseen jailbreak strategies. \mbox{SAA-850} transfers
released suffixes without re-optimization. Judge-based harm estimates are reported in the artifact and depend on automated harm evaluators; in a manual audit, only 3 of 40 HarmBench-positive completions satisfied a stricter actionable-harm criterion. The formatting-artifact mechanism is demonstrated only on \mbox{Llama-2}, and some cross-model detector results use adaptations of the released configurations, which limit the generalizability of these findings.


\section{Conclusion}
\label{sec:conc}

Our findings show that model-internal jailbreak detectors depend not only on ranking quality, but also on the data used to construct scores and thresholds. AUROC remains useful for measuring score separation, but it does not indicate protection at the low false-positive rates required in user-facing LLM systems. A more deployment-relevant metric is per-family TPR at a strict false-positive budget, together with realized FPR on deployment-like benign traffic. Calibration and reference data should be treated as part of the detector's security configuration. Calibration data should reflect realistic user traffic, while reference sets should be evaluated across multiple draws and audited for composition to identify unstable or uneven attack coverage. At scale, elevated false-positive rates can create unnecessary blocking and overhead, while low family-level detection can leave some attack families insufficiently covered. Reliable jailbreak defense thus requires evaluating the full data-dependent decision process rather than detector scores alone. The key question is whether diverse attacks are consistently detected under a false-positive budget a deployed system can sustain.

\section*{Acknowledgment}
Anthropic Claude was used to assist with writing and refining code used in the implementation and experimental workflow. All AI-assisted content and code were reviewed and verified by the authors, who take full responsibility for the final manuscript, implementation, and reported results.


\begin{thebibliography}{99}
\bibitem{zhou2025don}
Y. Zhou, J. Lou, Z. Huang, Z. Qin, S. Yang, and W. Wang,
``Don't Say No: Jailbreaking LLM by Suppressing Refusal,''
in \textit{Findings of the Association for Computational Linguistics: ACL 2025},
2025, pp. 25224--25249.
\bibitem{xu2024dpo}
S. Xu, W. Fu, J. Gao, W. Ye, W. Liu, Z. Mei, G. Wang, C. Yu, and Y. Wu,
``Is DPO Superior to PPO for LLM Alignment? A Comprehensive Study,''
arXiv:2404.10719, 2024.
\bibitem{qi2024visual}
X. Qi, K. Huang, A. Panda, P. Henderson, M. Wang, and P. Mittal,
``Visual Adversarial Examples Jailbreak Aligned Large Language Models,''
in \textit{Proc. AAAI Conf. Artificial Intelligence},
vol. 38, no. 19, 2024, pp. 21527--21536.
\bibitem{mtk} H.~Zhang \textit{et al.}, ``Defending jailbreak attacks on large language models via manifold trajectory kinetics,''
in \textit{Proc. USENIX Security Symposium}, 2026.
\bibitem{gradsafe} Y.~Xie, M.~Fang, R.~Pi, and N.~Gong, ``GradSafe: Detecting jailbreak prompts for LLMs
via safety-critical gradient analysis,'' in \textit{Proc. ACL}, 2024, pp.~507--518.
\bibitem{hiddendetect} Y.~Jiang \textit{et al.}, ``HiddenDetect:
Detecting jailbreak attacks against multimodal large language models via monitoring hidden states,'' in
\textit{Proc. ACL}, 2025, pp.~14880--14893.
\bibitem{tibshirani} R.~J. Tibshirani, R.~Foygel~Barber, E.~J. Cand\`es, and A.~Ramdas, ``Conformal prediction
under covariate shift,'' in \textit{Proc. NeurIPS}, 2019, pp.~2526--2536.
\bibitem{bradley} A.~P. Bradley, ``The use of the area under the ROC curve in the evaluation of machine
learning algorithms,'' \textit{Pattern Recognition}, vol.~30, no.~7, pp.~1145--1159, 1997.
\bibitem{mcclish} D.~K. McClish, ``Analyzing a portion of the ROC curve,'' \textit{Medical Decision
Making}, vol.~9, no.~3, pp.~190--195, 1989.
\bibitem{carlinimia} N.~Carlini, S.~Chien, M.~Nasr, S.~Song, A.~Terzis, and F.~Tram\`er, ``Membership
inference attacks from first principles,'' in \textit{Proc. IEEE Symposium on Security and Privacy}, 2022,
pp.~1897--1914.
\bibitem{iforest} F.~T. Liu, K.~M. Ting, and Z.-H. Zhou, ``Isolation forest,'' in \textit{Proc. IEEE
ICDM}, 2008, pp. 413--422.
\bibitem{dolly} M.~Conover \textit{et al.}, ``Free Dolly: Introducing the world's first truly open
instruction-tuned LLM,'' Databricks blog, 2023.
\bibitem{alpaca} R.~Taori \textit{et al.}, ``Stanford Alpaca: An instruction-following LLaMA model,'' GitHub
repository, 2023.
\bibitem{orbench} J.~Cui, W.-L. Chiang, I.~Stoica, and C.-J. Hsieh, ``OR-Bench: An over-refusal benchmark for
large language models,'' in \textit{Proc. ICML}, 2025, pp.~11515--11542.
\bibitem{advbench} A.~Zou, Z.~Wang, N.~Carlini, M.~Nasr, J.~Z. Kolter, and M.~Fredrikson, ``Universal
and transferable adversarial attacks on aligned language models,'' arXiv:2307.15043, 2023.
\bibitem{hexphi}
X.~Qi, Y.~Zeng, T.~Xie, P.-Y.~Chen, R.~Jia, P.~Mittal, and P.~Henderson,
``Fine-tuning aligned language models compromises safety, even when users do not intend to!,''
in \textit{Proc. ICLR}, 2024.
\bibitem{maliciousinstruct} Y.~Huang, S.~Gupta, M.~Xia, K.~Li, and D.~Chen, ``Catastrophic jailbreak of
open-source LLMs via exploiting generation,'' in \textit{Proc. ICLR}, 2024.
\bibitem{pku} J.~Ji \textit{et al.}, ``PKU-SafeRLHF: Towards multi-level safety alignment for LLMs with human
preference,'' in \textit{Proc. ACL}, 2025, pp.~31983--32016.
\bibitem{alon} G.~Alon and M.~Kamfonas, ``Detecting language model attacks with perplexity,''
arXiv:2308.14132, 2023.
\bibitem{jain} N.~Jain \textit{et al.}, ``Baseline defenses for adversarial attacks against aligned
language models,'' arXiv:2309.00614, 2023.
\bibitem{nasr} M.~Nasr \textit{et al.}, ``The attacker moves second: Stronger adaptive attacks bypass
defenses against LLM jailbreaks and prompt injections,'' in \textit{Proc. USENIX Security Symposium}, 2026.
\bibitem{bailey} L.~Bailey \textit{et al.}, ``Obfuscated activations bypass LLM latent-space defenses,''
in \textit{Proc. ICLR}, 2026.
\bibitem{llama2} H.~Touvron \textit{et al.}, ``Llama 2: Open foundation and fine-tuned chat models,''
arXiv:2307.09288, 2023.
\bibitem{llama3} A.~Grattafiori \textit{et al.}, ``The Llama 3 herd of models,'' arXiv:2407.21783, 2024.
\bibitem{mistral} A.~Q. Jiang \textit{et al.}, ``Mistral 7B,'' arXiv:2310.06825, 2023.
\bibitem{vicuna} L.~Zheng \textit{et al.}, ``Judging LLM-as-a-judge with MT-Bench and Chatbot Arena,'' in
\textit{NeurIPS Datasets and Benchmarks Track}, 2023.
\bibitem{harmbench} M.~Mazeika \textit{et al.}, ``HarmBench: A standardized evaluation framework for
automated red teaming and robust refusal,'' in \textit{Proc. ICML}, 2024.
\bibitem{strongreject} A.~Souly, Q.~Lu, D.~Bowen \textit{et al.}, ``A StrongREJECT for empty jailbreaks,''
in \textit{NeurIPS Datasets and Benchmarks Track}, 2024.
\bibitem{saa} M.~Andriushchenko, F.~Croce, and N.~Flammarion, ``Jailbreaking leading safety-aligned LLMs
with simple adaptive attacks,'' in \textit{Proc. ICLR}, 2025.
\bibitem{zulu} Z.-X. Yong, C.~Menghini, and S.~H. Bach, ``Low-resource languages jailbreak GPT-4,''
arXiv:2310.02446, 2023.
\bibitem{autodan} X.~Liu, N.~Xu, M.~Chen, and C.~Xiao, ``AutoDAN: Generating stealthy jailbreak prompts on aligned
large language models,'' in \textit{Proc. ICLR}, 2024.
\bibitem{drattack} X.~Li, R.~Wang, M.~Cheng, T.~Zhou, and C.-J. Hsieh, ``DrAttack: Prompt decomposition and
reconstruction makes powerful LLM jailbreakers,'' in \textit{Findings of EMNLP}, 2024, pp.~13891--13913.
\bibitem{ijp} X.~Shen, Z.~Chen, M.~Backes, Y.~Shen, and Y.~Zhang, ```Do Anything Now': Characterizing and
evaluating in-the-wild jailbreak prompts on large language models,'' in \textit{Proc. ACM CCS}, 2024,
pp.~1671--1685.
\bibitem{jailjudge} F.~Liu \textit{et al.}, ``JailJudge: A comprehensive
jailbreak judge benchmark with multi-agent enhanced explanation evaluation framework,'' arXiv:2410.12855, 2024.
\bibitem{pair} P.~Chao, A.~Robey, E.~Dobriban, H.~Hassani, G.~J. Pappas, and E.~Wong, ``Jailbreaking black box
large language models in twenty queries,'' in \textit{Proc. IEEE SaTML}, 2025, pp.~23--42.
\bibitem{tap} A.~Mehrotra \textit{et al.},
``Tree of attacks: Jailbreaking black-box LLMs automatically,'' in \textit{Proc. NeurIPS}, 2024.
\bibitem{pap} Y.~Zeng, H.~Lin, J.~Zhang, D.~Yang, R.~Jia, and W.~Shi, ``How Johnny can persuade LLMs to jailbreak
them: Rethinking persuasion to challenge AI safety by humanizing LLMs,'' in \textit{Proc. ACL}, 2024,
pp.~14322--14350.
\bibitem{toxicchat} Z.~Lin \textit{et al.}, ``ToxicChat: Unveiling hidden
challenges of toxicity detection in real-world user-AI conversation,'' in \textit{Findings of EMNLP}, 2023,
pp.~4694--4702.
\bibitem{lmsyschat} L.~Zheng \textit{et al.}, ``LMSYS-Chat-1M: A large-scale real-world LLM conversation
dataset,'' in \textit{Proc. ICLR}, 2024.
\bibitem{wildchat} W.~Zhao, X.~Ren, J.~Hessel, C.~Cardie, Y.~Choi, and Y.~Deng, ``WildChat: 1M ChatGPT interaction
logs in the wild,'' in \textit{Proc. ICLR}, 2024.
\bibitem{goldowsky} N.~Goldowsky-Dill, B.~Chughtai, S.~Heimersheim, and M.~Hobbhahn, ``Detecting strategic
deception with linear probes,'' in \textit{Proc. ICML}, 2025, pp.~19755--19786.
\bibitem{ccpp} H.~Cunningham \textit{et al.}, ``Constitutional classifiers++: Efficient production-grade
defenses against universal jailbreaks,'' arXiv:2601.04603, 2026.
\bibitem{axelsson} S.~Axelsson, ``The base-rate fallacy and the difficulty of intrusion detection,''
\textit{ACM Trans. Inf. Syst. Secur.}, vol.~3, no.~3, pp.~186--205, 2000.
\bibitem{sommer} R.~Sommer and V.~Paxson, ``Outside the closed world: On using machine learning for network
intrusion detection,'' in \textit{Proc. IEEE Symposium on Security and Privacy}, 2010, pp.~305--316.
\bibitem{arp} D.~Arp \textit{et al.}, ``Dos and don'ts of machine learning in computer security,'' in
\textit{Proc. USENIX Security Symposium}, 2022, pp.~3971--3988.
\bibitem{tesseract} F.~Pendlebury, F.~Pierazzi, R.~Jordaney, J.~Kinder, and L.~Cavallaro, ``TESSERACT: Eliminating
experimental bias in malware classification across space and time,'' in \textit{Proc. USENIX Security Symposium},
2019, pp.~729--746.
\bibitem{transcend} R.~Jordaney \textit{et al.}, ``Transcend: Detecting concept drift in malware classification
models,'' in \textit{Proc. USENIX Security Symposium}, 2017, pp.~625--642.
\bibitem{ovadia} Y.~Ovadia \textit{et al.}, ``Can you trust your model's uncertainty? Evaluating predictive
uncertainty under dataset shift,'' in \textit{Proc. NeurIPS}, 2019, pp.~13969--13980.
\bibitem{odin} S.~Liang, Y.~Li, and R.~Srikant, ``Enhancing the reliability of out-of-distribution image detection in
neural networks,'' in \textit{Proc. ICLR}, 2018.
\bibitem{bates} S.~Bates, E.~Cand\`es, L.~Lei, Y.~Romano, and M.~Sesia, ``Testing for outliers with conformal
p-values,'' \textit{Annals of Statistics}, vol.~51, no.~1, pp.~149--178, 2023.
\end{thebibliography}
\end{document}